# Three-dimensional non-Bosonic non-Fermionic quasiparticle through a quantized topological defect of crystal dislocation

Mingda Li[1], Qichen Song[1], M. S. Dresselhaus[2] and Gang Chen[1]

[1]Department of Mechanical Engineering, MIT, Cambridge, MA 02139, USA

[2]Department of Physics and Department of Electrical Engineering and Computer Sciences, MIT, Cambridge, MA 02139, USA

**It is a fundamental postulate that quasiparticles in 3D space obey either Bosonic or Fermionic statistics, satisfying either canonical commutation or anti-commutation relation[1]. However, under certain constraints, such as the 2D dimensional constraint, canonical quantization algebra is allowed to break down[2], and quasiparticles can obey other statistics, such as anyonic statistics[3]. In this study, we show that dislons– the quasiparticles in 3D due to quantized displacement field of a dislocation– can also obey neither Bosonic nor Fermionic statistics due to the topological constraint of the dislocation. With this theory, an effective electron field theory based on the electron-dislon interaction is obtained, which consists of two types of interactions. One classical-type of interaction is reducible to the well-known deformation potential scattering, and the other quantum-type of interaction indicates an effective attraction between electrons. The role of dislocations in superconductivity is clarified as the competition between the classical and quantum interactions, showing excellent agreement with experiments.**

Quasiparticles are fundamental emergent phenomena and building blocks in condensed matter physics, where a strongly-interacting microscopic system behaves like a weakly-interacting renormalized system[4]. In three-dimensional (3D) space, quasiparticles can be categorized as either Bosons or Fermions with canonical formalism, where Bosonic quasiparticles (aka collective excitations) include phonons, magnons, excitons, plasmons, etc., and Fermionic quasiparticles include quasi-electrons, holes and polarons, etc[1]. The renaissance of new quasiparticle families, such as Weyl and Dirac semimetals[5, 6, 7], hourglass fermions[8] and nodal-chain metals[9], attract wide recent interest, yet the resulting excitations can still be characterized as fermions[10] (or Bosons). On the other hand, in fact, when a constraint exists, the canonical formalism can break down if the constraint is inconsistent with the canonical commutation relations[2]. For instance in particle physics, canonical quantization of quantum electrodynamics is inconsistent with the Coulomb Gauge condition[11], while in condensed matter physics, lowering the dimension as a spatial constraint forbids the free exchange of particles, resulting in a nontrivial phase factor when one particle circulates around the other, namely anyonic statistics[3].

In this study, we show that even the simplest 3D isotropic solid can also accommodate quasiparticles with neither purely Bosonic nor Fermionic behavior, due to the topological definition of a crystal dislocation $\oint_C d\mathbf{u} = -\mathbf{b}$, where $\mathbf{u}$ is the lattice displacement field vector, $\mathbf{b}$ is Burgers vector and $C$ is an arbitrary loop enclosing the dislocation line[12]. The resulting quasiparticle, "dislon", upon quantizing the displacement field $\mathbf{u}$, is shown to be composed of two half-Bosonic fields. This can be understood intuitively through a comparison with a Dirac monopole, which also has to be characterized by two fields (but two classical vector fields) due to its nontrivial topology (Figure 1a)[13]. To explore the significance of this quasiparticle, the electron-dislocation interaction is studied in the present work, where the electron effective Hamiltonian is obtained using a method inspired by the Faddeev-Popov gauge fixing approach[14, 15] to impose the dislocation's topological constraint. The effective electron Hamiltonian is shown to be composed of three terms- a diagonal quadratic term (non-interacting electron), an off-diagonal quadratic term (classical scattering) and a quartic term (quantum-mechanical interaction). The classical scattering term describes the electron-dislocation scattering process with electron momentum transfer perpendicular to dislocation direction, which can be reduced to the well-known deformation potential scattering under long wavelength limit. On the other hand, the quantum interaction describes an effective attraction between electron pairs mediated by a quantized dislocation (Figure 1b). The elusive role of the crystal dislocation on superconductivity is clarified using this approach, where a modified BCS gap equation incorporating both the classical and quantum interactions is derived, capable of quantitatively describing the influence of the dislocation on the superconducting transition temperature $T_c$. It turns out that the competition between the classical and quantum interactions plays the governing role in determining $T_c$. To validate the theory, the $T_c$ of as many as ten dislocated superconductors are compared and excellent agreement is obtained.

**The foundation of the quantized dislocation**

The best way to understand the quantized dislocation is probably to compare it with a phonon. A phonon is a quantized lattice wave which can be mode-expanded in terms of plane waves[16]:

$$\mathbf{u}^{ph}(\mathbf{R}) = \frac{1}{\sqrt{N}} \sum_{\mathbf{k}} u_{\mathbf{k}}^{ph} \boldsymbol{\varepsilon}_{\mathbf{k}\lambda} e^{i\mathbf{k}\cdot\mathbf{R}} \tag{1}$$

where $\mathbf{u}^{ph}$ is the lattice displacement at spatial position $\mathbf{R}$, $\mathbf{\varepsilon}_{\mathbf{k}\lambda}$ is the polarization vector and $u_{\mathbf{k}}^{ph}$ is the lattice displacement in the $\mathbf{k}$-component mode. In the static limit, there is no displacement, i.e. $u_{\mathbf{k}}^{ph} = 0$. As to dislocation, inspired by mode-expansion work in 1D[17, 18], here we expand the lattice displacement for a dislocation line as

$$\mathbf{u}_i(\mathbf{R}) = \frac{1}{A} \sum_{\mathbf{k} \equiv (\mathbf{s}, \kappa)} F_i(\mathbf{k}) e^{i\mathbf{k} \cdot \mathbf{R}} u_{\mathbf{k}} \qquad (2)$$

where $A$ is the sample area, $u_{\mathbf{k}}$ is dimensionless displacement, $F_i(\mathbf{k})$ is an expansion function, $\mathbf{s}$ denotes the 2D momentum perpendicular to the dislocation direction, and $\kappa$ is the dislocation-direction momentum introduced here for later convenience. In a 3D isotropic solid, for a dislocation line along the z-direction and the glide plane within xz-plane, $F_i(\mathbf{k})$ can be written explicitly as

$$F_i(\mathbf{k}) \equiv F_i(\mathbf{s}; \kappa) = + \frac{1}{k_x k^2} \left( n_i(\mathbf{b} \cdot \mathbf{k}) + b_i(\mathbf{n} \cdot \mathbf{k}) - \frac{1}{(1-\nu)} \frac{k_i(\mathbf{n} \cdot \mathbf{k})(\mathbf{b} \cdot \mathbf{k})}{k^2} \right) \qquad (3)$$

where $i = x, y, z$ are the spatial components, $\mathbf{n}$ is the glide plane normal direction, $\mathbf{b}$ is the Burgers vector, and $\nu$ is the Poisson ratio. The reason for the mode expansion in Eq. (2) is straightforward. Under the following boundary condition,

$$\lim_{\kappa \to 0} u_{\mathbf{k}} \equiv \lim_{\kappa \to 0} u_{\mathbf{s}, \kappa} = 1, \text{ for } \forall \mathbf{s} \qquad (4)$$

Eq. (2) becomes equivalent to textbook results of static dislocation, for both edge and screw dislocations (Supporting information I). In fact, Eq. (4) – the topological definition of the dislocation, i.e. the reducibility to a classical dislocation– can be considered as the essential starting point of this theory, leading to the breakdown of the canonical quantization.

A brief comparison of a phonon and a quantized dislocation (aka dislon) is summarized in Table 1.

### The dislocation Hamiltonian in 3D

Using the mode expansion Eq. (2), the classical dislocation Hamiltonian composed of kinetic energy $T$ and potential energy $U$ can be written in the usual way[19] as (Supporting information II)

$$H = T + U = \frac{\rho}{2} \int \sum_{i=1}^{3} \dot{\mathbf{u}}_i^2(\mathbf{R}) d^3\mathbf{R} + \frac{1}{2} \int c_{ijkl} u_{ij} u_{kl} d^3\mathbf{R} = \frac{1}{2L} \sum_{\mathbf{k}} T(\mathbf{k}) \dot{u}_{\mathbf{k}} \dot{u}_{-\mathbf{k}} + \frac{1}{2L} \sum_{\mathbf{k}} U(\mathbf{k}) u_{\mathbf{k}} u_{-\mathbf{k}} \qquad (5)$$

where $\rho$ is the mass density, $c_{ijkl}$ is the stiffness tensor which can be written as $c_{ijkl} = \lambda \delta_{ij}\delta_{kl} + \mu(\delta_{ik}\delta_{jl} + \delta_{il}\delta_{jk})$ with $\lambda$ the Lamé's first parameter and $\mu$ the shear modulus in an isotropic medium, and the coefficients $T(\mathbf{k}) \equiv \rho|F(\mathbf{k})|^2$, $U(\mathbf{k}) \equiv (\lambda+\mu)[\mathbf{k}\cdot\mathbf{F}(\mathbf{k})]^2 + \mu k^2|F(\mathbf{k})|^2$. By noticing that the conjugate momentum is defined as $p_\mathbf{k} = \frac{\partial L}{\partial \dot{u}_\mathbf{k}} = m_\mathbf{k}\dot{u}_{-\mathbf{k}}$, and by further defining $m_\mathbf{k} \equiv T(\mathbf{k})/L$, $\Omega_\mathbf{k} \equiv \sqrt{U(\mathbf{k})/T(\mathbf{k})}$, the classical dislocation Hamiltonian Eq. (5) can be rewritten in the following way

$$H = \sum_\mathbf{k} \frac{p_\mathbf{k} p_{-\mathbf{k}}}{2m_\mathbf{k}} + \sum_\mathbf{k} \frac{m_\mathbf{k}\Omega_\mathbf{k}^2}{2} u_\mathbf{k} u_{-\mathbf{k}} \tag{6}$$

This appears to be similar to the quantization of the phonon, but with a momentum-dependent mass term. However, there is huge difference due to the constraint Eq. (4). To see this, in order to obtain a quantized theory of Eq. (6), it is natural to introduce canonical quantization as the case of phonon

$$\begin{cases} u_\mathbf{k} = Z_\mathbf{k}\left[a_\mathbf{k} + a^+_{-\mathbf{k}}\right] \\ p_\mathbf{k} = \frac{i\hbar}{2Z_\mathbf{k}}\left[a^+_\mathbf{k} - a_{-\mathbf{k}}\right] \end{cases} \tag{7}$$

where $Z_\mathbf{k} \equiv \sqrt{\hbar/2m_\mathbf{k}\Omega_\mathbf{k}}$ is a prefactor. However, if we adopt the usual Bosonic commutation relation $[a_\mathbf{k}, a^+_{\mathbf{k}'}] = \delta_{\mathbf{k}\mathbf{k}'}$, it can be shown that Bosonic commutator is incompatible with the boundary condition in Eq. (4):

$$\lim_{\kappa \to 0}\delta_{\mathbf{k}\mathbf{k}'} = \lim_{\kappa \to 0}[a_{\mathbf{s},\kappa}, a^+_{\mathbf{s},\kappa}] = \lim_{\kappa \to 0}[\frac{1}{Z_\mathbf{k}} - a^+_{-\mathbf{s},-\kappa}, \frac{1}{Z_\mathbf{k}} - a_{-\mathbf{s},-\kappa}] = -\lim_{\kappa \to 0}[a_{-\mathbf{s},-\kappa}, a^+_{-\mathbf{s},-\kappa}] = -\lim_{\kappa \to 0}\delta_{\mathbf{k}\mathbf{k}'}$$

which leads to a contradiction. Instead, we find that the following statistics is fully consistent with the boundary condition in Eq. (4):

$$[a_\mathbf{k}, a^+_{\mathbf{k}'}] = \delta_{\mathbf{k}\mathbf{k}'}\,\text{sgn}(\mathbf{k}) \tag{8}$$

where the vector-sgn function is defined based on a generalization of complex-sgn function as

$$\text{sgn}(\mathbf{k}) = \begin{cases} +1, & \text{if } k_x > 0 \\ -1, & \text{if } k_x < 0 \\ \text{sgn}(k_y,\kappa), & \text{if } k_x = 0 \end{cases}, \text{sgn}(k_y,\kappa) = \begin{cases} 1, & \text{if } k_y > 0 \\ -1, & \text{if } k_y < 0 \\ \text{sgn}\,\kappa, & \text{if } k_y = 0 \end{cases}, \text{sgn}(\kappa) = \begin{cases} 1, & \text{if } \kappa > 0 \\ -1, & \text{if } \kappa < 0 \\ 0, & \text{if } \kappa = 0 \end{cases} \tag{9}$$

The quasiparticle in Eq. (8) fully defines a quantized dislocation, which we call a "dislon", while $\Omega_\mathbf{k}$ is understood as dislon dispersion. Now we have

$$\lim_{\kappa \to 0}\text{sgn}(\mathbf{k}) = \lim_{\kappa \to 0}[a_{s,\kappa}, a^+_{s,\kappa}] = \lim_{\kappa \to 0}[\frac{1}{Z_\mathbf{k}} - a^+_{-s,-\kappa}, \frac{1}{Z_\mathbf{k}} - a_{-s,-\kappa}] = -\lim_{\kappa \to 0}[a_{-s,-\kappa}, a^+_{-s,-\kappa}] = -\lim_{\kappa \to 0}\text{sgn}(-\mathbf{k}) = \lim_{\kappa \to 0}\text{sgn}(\mathbf{k})$$

showing perfect consistency between the constraint Eq. (4) and dislon algebra Eq. (8). A few useful identities relevant to Eq. (8) are listed in the Supporting Information II.

In other words, the topological constraint of the dislocation Eq. (4) prevents the quantized dislocation displacement from being purely Bosonic. In the following we prove that Eq. (8) actually indicates a Boson sea behavior composed to two half-Bosons. The concept of a Boson sea[20] is similar to a Fermi sea and has an origin from supersymmetry[21]: for fermions when wavevector $k$ above the Fermi sea $k > k_F$, a normal electron creation operator $c^+_{k>k_F}$ is defined, while below the Fermi sea $k < k_F$, a hole creation operator is defined as $d^+_{k<k_F} = c_{k<k_F}$. Here when $\mathbf{k}$ is above the Boson sea $\text{sgn}(\mathbf{k}) > 0$, we have dislon creation operator $a^+_{\mathbf{k}>0}$, while when below Boson sea $\text{sgn}(\mathbf{k}) < 0$, we define a new operator $b^+_{-\mathbf{k}} = a_\mathbf{k}$. In this way, we have $b_{-\mathbf{k}} = a^+_\mathbf{k}$ valid, and the operators $b_\mathbf{k}$ and $b^+_\mathbf{k}$ satisfies normal Bosonic algebra $[b_\mathbf{k}, b^+_{\mathbf{k}'}] = \delta_{\mathbf{kk}'}$. The normal canonical quantization commutator is thereinafter fully recovered

$$[a_\mathbf{k}, a^+_{\mathbf{k}'}] = \delta_{\mathbf{kk}'}, \text{ for } \text{sgn}(\mathbf{k})>0$$
$$[b_\mathbf{k}, b^+_{\mathbf{k}'}] = \delta_{\mathbf{kk}'}, \text{ for } \text{sgn}(\mathbf{k})<0 \tag{10}$$

The quantized dislocation Hamiltonian Eq. (6) can be written as (see Supporting information II)

$$H = \sum_\mathbf{k} \hbar\Omega(\mathbf{k})a^+_\mathbf{k} a_\mathbf{k} = \sum_{\mathbf{k}\geq 0}\hbar\Omega(\mathbf{k})\left(a^+_\mathbf{k} a_\mathbf{k} + \frac{1}{2}\right) + \sum_{\mathbf{k}\geq 0}\hbar\Omega(\mathbf{k})\left(b^+_\mathbf{k} b_\mathbf{k} + \frac{1}{2}\right) \tag{11}$$

where we have used the fact that $\Omega(\mathbf{k}=\mathbf{0}) = 0$, and the short-hand notation $\mathbf{k} \geq 0$ denoting $\text{sgn}\,\mathbf{k} \geq 0$. The necessity of two Boson fields (Figure 1b) in representing a quantized dislocation resembles a Dirac monopole[13], where two magnetic vector potentials have to be defined for the south and north poles, respectively (Figure 1a)- in both scenarios, due to the topological constraint, a single field quantity – whether a classical vector potential or a quantized Bosonic field – is simply not sufficient to describe the intrinsic topological feature.

**The electron-dislon interaction**

The generic electron-ion interacting Hamiltonian can be represented using the deformation potential approximation as[16]

$$H_{e-ion} = \int d^3\mathbf{R}\rho_e(\mathbf{R})\sum_{j=1}^{N}\nabla_{\mathbf{R}}V_{ei}(\mathbf{R}-\mathbf{R}_j^0)\cdot\mathbf{u}_j \qquad (12)$$

where $\rho_e(\mathbf{R})$ is the charge density operator which is defined as $\rho_e(\mathbf{R}) = en_e(\mathbf{R}) = \frac{e}{V}\sum_{\mathbf{kp}\sigma}e^{-i\mathbf{p}\cdot\mathbf{R}}c^+_{\mathbf{k+p}\sigma}c_{\mathbf{k}\sigma}$, the Coulomb potential $V_{ei}(\mathbf{R}-\mathbf{R}_j^0) = \frac{1}{V}\sum_{\mathbf{k}}V_{\mathbf{k}}e^{i\mathbf{k}\cdot(\mathbf{R}-\mathbf{R}_j^0)}$ has a Fourier component $V_{\mathbf{k}} \equiv \frac{4\pi Ze}{k^2+k_{TF}^2}$, in which $k_{TF}$ is the Thomas-Fermi screening wavenumber. Using Eq. (2), the electron-quantized dislocation (aka dislon) interacting Hamiltonian Eq. (12) can be further simplified as ( see Supporting information III)

$$\begin{aligned}H_{e-dis} &= \sum_{\mathbf{k'k}\sigma} g_{\mathbf{k}}c^+_{\mathbf{k'+k}\sigma}c_{\mathbf{k'}\sigma}(a_{\mathbf{k}}+a^+_{-\mathbf{k}}) \\ &= \sum_{\substack{\mathbf{k'}\sigma \\ \mathbf{k}\geq 0}} g_{\mathbf{k}}c^+_{\mathbf{k'+k}\sigma}c_{\mathbf{k'}\sigma}(a_{\mathbf{k}}+b_{\mathbf{k}}) + \sum_{\substack{\mathbf{k'}\sigma \\ \mathbf{k}\geq 0}} g^*_{\mathbf{k}}c^+_{\mathbf{k'-k}\sigma}c_{\mathbf{k'}\sigma}(b^+_{\mathbf{k}}+a^+_{\mathbf{k}})\end{aligned} \qquad (13)$$

with the electron-dislocation coupling constant $g_{\mathbf{k}} \equiv \frac{eN}{VA}V_{\mathbf{k}}[i\mathbf{k}\cdot\mathbf{F}(\mathbf{k})]\sqrt{\frac{\hbar}{2m_{\mathbf{k}}\Omega_{\mathbf{k}}}}$, and $N$ is the number of ions in the system, and $V$ is the sample volume. As a comparison, the electron-phonon interaction Hamiltonian is written as[16]

$$H_{e-ph} = \sum_{\mathbf{k'k}\sigma} g^{ph}_{\mathbf{k}}c^+_{\mathbf{k'+k}\sigma}c_{\mathbf{k'}\sigma}(A_{\mathbf{k}}+A^+_{-\mathbf{k}}) \qquad (14)$$

with the electron-phonon coupling constant $g^{ph}_{\mathbf{k}} = \frac{ieV_{\mathbf{k}}}{V}(\mathbf{k}\cdot\varepsilon_{\mathbf{k}})\sqrt{\frac{N\hbar}{2M\omega_{\mathbf{k}}}}$. There are two major differences between the electron-dislon interaction and the electron-phonon interaction

1. $a_{\mathbf{k}}$ for dislons are not purely Bosonic, but satisfy Eq. (8); for phonons, the operators $A_{\mathbf{k}}$ are Bosonic.

2. $g_{\mathbf{k}}$ and $g^{ph}_{\mathbf{k}}$ have different levels of extensivity, i.e. the dependence of scale $L$: $g^{ph}_{\mathbf{k}} \sim \frac{1}{L^{3/2}}$ while $g_{\mathbf{k}} \sim \frac{1}{L^2}$. This is quite reasonable since up to now we are studying a single dislocation only. For multiple independent dislocation, similar to the phonon case multiply $\sqrt{N}$ where $N$ is number of ions, here, we multiply $g_\mathbf{k}$ by a factor of $\sqrt{N_D}$, where $N_D$ is the number of dislocation. To do so, we recover the same scale dependence $g_\mathbf{k} \sim \dfrac{1}{L^{3/2}}$ as with a phonon. The factor of $\sqrt{N_D}$ is also necessary to correctly reproduce the classical relaxation rate (See below).

Therefore, for multiple independent dislocations, we have the electron-dislocation coupling constant

$$g_\mathbf{k} = \frac{eN}{VA}\sqrt{N_D} V_\mathbf{k}[i\mathbf{k}\cdot\mathbf{F}(\mathbf{k})]\sqrt{\frac{\hbar}{2m_\mathbf{k}\Omega_\mathbf{k}}} \tag{15}$$

## Effective field theory of electrons

To see how electron-dislocation interaction Eq. (13) under the constraint of Eq. (4) really affects the material electronic properties, we adopt a functional integral approach[22] to eliminate the dislon degree of freedom and establish an effective field theory for solely electrons. The total action of the interacting electron-dislon system can be written as

$$S_{tot}[\bar{\psi},\psi,\bar{\chi},\chi] = S_e[\bar{\psi},\psi] + S_{dis}[\bar{\chi}_a,\chi_a,\bar{\chi}_b,\chi_b] + S_{e-dis}[\bar{\psi},\psi,\bar{\chi}_a,\chi_a,\bar{\chi}_b,\chi_b] \tag{16}$$

where $\bar{\psi},\psi$ are electron fields and $\bar{\chi},\chi$ are dislon fields, with $a$ and $b$ are the 2 Bosonic components of a dislon. Now the non-interacting electron and dislon actions can be written as

$$\begin{aligned}
S_e[\bar{\psi},\psi] &= \sum_{n\mathbf{p}\sigma}\bar{\psi}_{n\mathbf{k}\sigma}(-ip_n + \varepsilon_\mathbf{k} - \mu)\psi_{n\mathbf{k}\sigma} \\
S_{dis}[\bar{\chi},\chi] &= \sum_{n,\mathbf{k}\geq 0}\left[\bar{\chi}_{a\mathbf{k}n}(-i\omega_n + \hbar\Omega_\mathbf{k})\chi_{a\mathbf{k}n} + \bar{\chi}_{b\mathbf{k}n}(-i\omega_n + \hbar\Omega_\mathbf{k})\chi_{b\mathbf{k}n}\right]
\end{aligned} \tag{17}$$

where $\omega_n \equiv 2n\pi T$, $p_n \equiv (2n+1)\pi T$ are Bosonic and Fermionic Matsubara frequencies, respectively. The interaction Hamiltonian Eq. (13) can be also rewritten in action form as

$$S_{e-dis}[\bar{\psi},\psi,\bar{\chi},\chi] = \sum_{n,\mathbf{k}\geq 0}\frac{g_\mathbf{k}}{\sqrt{\beta}}\rho_{n\mathbf{k}}(\chi_{n\mathbf{k},a} + \chi_{n\mathbf{k},b}) + \sum_{n,\mathbf{k}\geq 0}\frac{g_\mathbf{k}^*}{\sqrt{\beta}}\rho_{-n-\mathbf{k}}(\bar{\chi}_{n\mathbf{k},a} + \bar{\chi}_{n\mathbf{k},b}) \tag{18}$$

where we have defined the charge density in momentum space as $\rho_{n\mathbf{k}} \equiv \sum_{\mathbf{k}'m\sigma}\bar{\psi}_{m+n,\mathbf{k}'+\mathbf{k}\sigma}\psi_{m\mathbf{k}'\sigma}$. To achieve further simplification, we perform the Keldysh rotation[23], by defining that

$$d_{\mathbf{k}n} = \frac{1}{\sqrt{2}}\left(\chi_{\mathbf{k}n,a} + \chi_{\mathbf{k}n,b}\right), \quad \bar{d}_{\mathbf{k}n} = \frac{1}{\sqrt{2}}\left(\bar{\chi}_{\mathbf{k}n,a} + \bar{\chi}_{\mathbf{k}n,b}\right)$$
$$f_{\mathbf{k}n} = \frac{1}{\sqrt{2}}\left(-\chi_{\mathbf{k}n,a} + \chi_{\mathbf{k}n,b}\right), \quad \bar{f}_{\mathbf{k}n} = \frac{1}{\sqrt{2}}\left(-\bar{\chi}_{\mathbf{k}n,a} + \bar{\chi}_{\mathbf{k}n,b}\right)$$

(19)

Then the non-interacting dislon action in Eq. (17) and interaction action Eq. (18) can be rewritten as

$$S_{dis} = \sum_{n,\mathbf{k}\geq 0} \bar{d}_{\mathbf{k}n}(-i\omega_n + \hbar\Omega_{\mathbf{k}})d_{\mathbf{k}n} + \sum_{n,\mathbf{k}\geq 0} \bar{f}_{\mathbf{k}n}(-i\omega_n + \hbar\Omega_{\mathbf{k}})f_{\mathbf{k}n}$$

$$S_{e-dis}[\bar{\psi},\psi,\bar{d},d] = \sum_{n,\mathbf{k}\geq 0} \sqrt{\frac{2}{\beta}} g_{\mathbf{k}} \rho_{n\mathbf{k}} d_{n\mathbf{k}} + \sum_{n,\mathbf{k}\geq 0} \sqrt{\frac{2}{\beta}} g_{\mathbf{k}}^* \rho_{-n-\mathbf{k}} \bar{d}_{n\mathbf{k}}$$

(20)

respectively. Now since the $f_{\mathbf{k}n}$ field from Eq. (20) is completely decoupled from the electron degree of freedom, it can be neglected when defining the electron effective action. The constraint in Eq. (4) can now be rewritten as

$$\lim_{\kappa\to 0} d_{\mathbf{s},\kappa} = d_{\mathbf{s}0} = \lim_{\kappa\to 0}\sqrt{\frac{\beta m_{\mathbf{k}}\Omega_{\mathbf{k}}}{\hbar}} \equiv C_{\mathbf{s}}$$
$$\lim_{\kappa\to 0} \bar{d}_{\mathbf{s},\kappa} = \bar{d}_{\mathbf{s}0} = \lim_{\kappa\to 0}\sqrt{\frac{\beta m_{\mathbf{k}}\Omega_{\mathbf{k}}}{\hbar}} = C_{\mathbf{s}}$$

(21)

where $C_{\mathbf{s}} = \lim_{\kappa\to 0}\sqrt{\frac{\beta m_{\mathbf{k}}\Omega_{\mathbf{k}}}{\hbar}}$ is an $\mathbf{s}$-dependent constant satisfying $\text{sgn}(\mathbf{s})\geq 0$ since the original single displacement $u_{\mathbf{s}0}$ is divided into $d_{\mathbf{s}0}$ and $\bar{d}_{\mathbf{s}0}$. With Eq. (20) at hand, the effective electron action can be defined in analog to the Faddeev-Popov gauge fixing method[15] to impose the constraint Eq. (21) using δ-functions as

$$\exp(-S_{e\!f\!f}[\bar{\psi},\psi]) \equiv \exp(-S_e[\bar{\psi},\psi])\times \int \begin{array}{c} D[\bar{d},d]\prod\limits_{n\mathbf{s}\geq 0}\delta(d_{n\mathbf{s}0} - C_{\mathbf{s}})\delta(\bar{d}_{n\mathbf{s}0} - C_{\mathbf{s}})\times \\ \exp(-S_{dis}[\bar{d},d] - S_{e-dis}[\bar{\psi},\psi,\bar{d},d]) \end{array}$$

(22)

After a few functional integration steps (Supporting Information IV), we obtain the effective electron action upon the electron-dislocation interaction

$$S_{e\!f\!f}[\bar{\psi},\psi] = S_e[\bar{\psi},\psi] - \sum_{n\mathbf{k}|\kappa\neq 0}\frac{\hbar\Omega_{\mathbf{k}} g_{\mathbf{k}}^* g_{\mathbf{k}}}{\beta(\omega_n^2 + \hbar^2\Omega_{\mathbf{k}}^2)}\rho_{-n-\mathbf{k}}\rho_{n\mathbf{k}} + \sum_{\mathbf{k}n\sigma}\sum_{m,\mathbf{s}}\frac{C_{\mathbf{s}}}{\sqrt{2\beta}}\left(g_{\mathbf{s}}\bar{\psi}_{n+m\mathbf{k}+\mathbf{s}\sigma} + g_{\mathbf{s}}^*\bar{\psi}_{n-m\mathbf{k}-\mathbf{s}\sigma}\right)\psi_{n\mathbf{k}\sigma}$$

(23)

Noticing that in real materials, dislocations can arise in many directions, and moreover since the $\kappa\neq 0$ constraint in Eq. (23) only contains a small volume of phase space (which is negligible in the thermodynamic limit), we can safely relax this constraint. Now if performing an analytical continuation step back to the real frequency and picking-up the Cooper channel in the quartic term, the effective electron Hamiltonian is finally written as a sum of three contributions, the non-interacting electron $H_0$, and classical scattering $H_c$ and quantum interaction $H_q$,

$$H_{eff} = H_0 + H_c + H_q$$
$$= \sum_{k\sigma}(\varepsilon_k - \mu)c^+_{k\sigma}c_{k\sigma} + \sum_{k\sigma}\sum_s \left(A_s c^+_{k+s\sigma} + A^*_s c^+_{k-s\sigma}\right)c_{k\sigma} + \sum_{qkk'} V_{eff}(q) c^+_{k+q\uparrow} c^+_{-k\downarrow} c_{-k'+q\downarrow} c_{k'\downarrow} \quad (24)$$

in which the classical electron-dislocation scattering (Figure 1b green straight line) amplitude $A_s$ is defined as

$$A_s = \frac{eN}{2VA}\sqrt{N_{dis}} V_s [i\mathbf{s}\cdot\mathbf{F}(\mathbf{s})] = \frac{ieN}{2VL}\sqrt{n_{dis}} V_s \left(\frac{1-2\nu}{1-\nu}\right)\frac{(\mathbf{n}\cdot\mathbf{s})(\mathbf{b}\cdot\mathbf{s})}{k_x s^2} \quad (25)$$

where $s \equiv \sqrt{k_x^2 + k_y^2}$. The quantum-mechanical electron-electron attractive interaction mediated by the retarded dislon (Figure 1b wavy line) is written as

$$V_{eff}(\mathbf{q}) = -\frac{\hbar\Omega_q g^*_q g_q}{-\omega^2 + \hbar^2\Omega_q^2} \sim -\frac{g^*_q g_q}{\hbar\Omega_q} = -\left(\frac{N}{VA}\right)^2 N_{dis}\frac{(eV_q)^2}{2m_q\Omega_q^2}[\mathbf{q}\cdot F(\mathbf{q})]^2 \quad (26)$$

The ~ approximation is valid since electrons near Fermi surface can have very small energy transfer ω.

Eqs. (24)- (26) are the main result of electron Hamiltonian, showing how dislocations will interact with electrons from a quantitative many-body viewpoint. Interestingly, the effective attractive interaction mediated by the dislon Eq. (26) shares some structural similarity with the interaction mediated by the phonon[24], which leads to phonon-mediated superconductivity. Here the interaction has a different coupling constant and the phonon dispersion $\omega_q$ is replaced by dislon dispersion $\Omega_q$, as compared in Table 1.

**Classical electron-dislocation scattering**

Qualitatively, the classical scattering term $H_c$ is quite intuitive in describing an electron scattering process $\mathbf{k} \to \mathbf{k}+\mathbf{s}$ and $\mathbf{k} \to \mathbf{k}-\mathbf{s}$, with momentum changes only happening perpendicular to the dislocation direction with amplitude $A_s$ and $A^*_s$, respectively (Figure 1b). Quantitatively, for an edge dislocation $\mathbf{b} = (b\ 0\ 0)$, $\mathbf{b}\cdot\mathbf{s} = bk_x$, $\mathbf{n}\cdot\mathbf{s} = s\sin\theta$, the Fourier transform of the scattering amplitude gives (Supporting information V)

$$A^{Edge}(\mathbf{r}) = \int A_\mathbf{s}^{Edge} e^{i\mathbf{s}\cdot\mathbf{r}} \frac{d^2\mathbf{s}}{4\pi^2} = \frac{N}{VL} \frac{2\pi Ze^2}{k_{TF}^2} \times \frac{b}{2\pi}\left(\frac{1-2\nu}{1-\nu}\right)\frac{\sin\theta}{r} \propto \frac{b}{2\pi}\left(\frac{1-2\nu}{1-\nu}\right)\frac{\sin\theta}{r} \quad (27)$$

Aside from a proportionality constant, Eq. (27) is in perfect agreement with the classical dislocation deformation scattering potential[25], where $\delta V = -\frac{a_c b}{2\pi}\left(\frac{1-2\nu}{1-\nu}\right)\frac{\sin\theta}{r}$ with $a_c$ is the conduction band deformation potential which is an empirical parameter and has a similar order of magnitude with the prefactor in Eq. (27). For a screw dislocation in an isotropic crystal, it is well-known that only purely shear strain exists, causing no dilation or compression of the unit cells hence leading to no deformation potential scattering[25]. This is consistent with the result in Eq. (25), that for a screw dislocation $\mathbf{b}=(0\ 0\ b)$, we have $A_\mathbf{s}^{Screw}=0$.

The classical scattering term $H_c$, despite being quadratic, is full of off-diagonal elements which complicates the computation. To further simplify the classical scattering terms, from physical grounds in the following we consider elastic dislocation-electron scattering only, so that the relaxation rate $\Gamma_\mathbf{k}$ can be computed using Fermi's Golden rule as

$$\Gamma_\mathbf{k} \equiv \frac{\hbar}{2\tau_\mathbf{k}} = \frac{\pi}{2}\sum_\mathbf{s}|\langle\mathbf{k}+\mathbf{s}|\hat{A}_c|\mathbf{k}\rangle|^2 \delta(\varepsilon_{\mathbf{k}+\mathbf{s}}-\varepsilon_\mathbf{k}) \quad (28)$$

where the sum is over all final states which transfer in-plane momentum but maintaining elastic scattering, and $\hat{A}_c$ is the classical scattering amplitude operator. The relaxation rate can thus be computed explicitly as (Supporting Information V)

$$\Gamma_\mathbf{k} \sim \frac{\pi m^*}{4\hbar^2 k^2 k_{TF}^4}\left(\frac{Ze^2 N}{V}\right)^2 n_{dis} b^2 \left(\frac{1-2\nu}{1-\nu}\right)^2 \quad (29)$$

where $m^*$ is electron effective mass. With the relaxation rate at hand, the original quadratic Hamiltonian in Eq. (24) can be approximated by a non-Hermitian Hamiltonian [26] with complex eigenvalues as

$$\begin{aligned}H_0 + H_c &= \sum_{\mathbf{k}\sigma}(\varepsilon_\mathbf{k}-\mu)c_{\mathbf{k}\sigma}^+ c_{\mathbf{k}\sigma} + \sum_{\mathbf{k}\sigma}\sum_\mathbf{s}\left(A_\mathbf{s} c_{\mathbf{k}+\mathbf{s}\sigma}^+ + A_\mathbf{s}^* c_{\mathbf{k}-\mathbf{s}\sigma}^+\right)c_{\mathbf{k}\sigma} \\ &\approx \sum_\mathbf{k}(E_\mathbf{k}-\mu'+i\Gamma_\mathbf{k})c_{\mathbf{k}\uparrow}^+ c_{\mathbf{k}\uparrow} + \sum_\mathbf{k}(E_\mathbf{k}-\mu'-i\Gamma_\mathbf{k})c_{\mathbf{k}\downarrow}^+ c_{\mathbf{k}\downarrow}\end{aligned} \quad (30)$$

The reason for this definition is to preserve time-reversal symmetry which the original Hamiltonian processes[26], that $T(H_0+H_c)T^{-1} = H_0+H_c$. Meanwhile, since electron-dislocation scattering is expected to increase the electron effective mass and decrease the energy[18], with the same order as $\Gamma_\mathbf{k}$, for simplicity we assume $E_\mathbf{k}-\mu' \approx \varepsilon_\mathbf{k}-\Gamma_\mathbf{k}-\mu$ valid.

### Electron-dislon BCS superconductivity

With simplified diagonal quadratic Hamiltonian Eq. (30), we are now ready to study the effect of dislocation on superconductivity at a quantitative level. Unlike the well-understood role of dislocation in affecting superconducting critical magnetic field $H_c$ due to flux-pinning mechanism[27], the role of dislocation to superconducting transition temperature $T_c$ remains elusive. For instance, in superconducting tantalum, the decrease of superconducting transition temperature $T_c$ is correlated with the decrease of the electron mean free path, while equal decreases of the mean free path by dislocation has no effect on $T_c$[28]. Anderson proposed a mechanism[29] of anisotropy which is capable of increasing $T_c$, however for the particular case of the dislocation, no quantitative agreement has been achieved. Here, under the local-coupling limit[30] (i.e. $V_{eff}(\mathbf{q})$ has no $\mathbf{q}$-dependence), the electron Hamiltonian Eq. (24) can be written as

$$H_{eff} = \sum_{\mathbf{k}}\left[(E_{\mathbf{k}} - \mu' + i\Gamma_{\mathbf{k}})c^+_{\mathbf{k}\uparrow}c_{\mathbf{k}\uparrow} + (E_{\mathbf{k}} - \mu' - i\Gamma_{\mathbf{k}})c^+_{\mathbf{k}\downarrow}c_{\mathbf{k}\downarrow}\right] - \frac{g_{dis}}{V}\sum_{\mathbf{qkk'}}c^+_{\mathbf{k+q}\uparrow}c^+_{-\mathbf{k}\downarrow}c_{-\mathbf{k'+q}\downarrow}c_{\mathbf{k'}\uparrow} \quad (31)$$

where

$$g_{dis} = -\langle V_{eff}(\mathbf{q})\rangle V = \left(\frac{1-2\nu}{1-\nu}\right)^2 b^2 \left(\frac{N}{V}\right)^2 n_{dis}L\left\langle \frac{(eV_{\mathbf{q}})^2}{2m_{\mathbf{q}}\Omega_{\mathbf{q}}^2} \frac{(\mathbf{n}\cdot\mathbf{q})^2(\mathbf{b}\cdot\mathbf{q})^2}{q_x^2 q^4 b^2}\right\rangle \quad (32)$$

Now if we add the contribution from the phonon-induced electron attraction contribution $g_{ph}$, Eq. (31) can be further simplified as (Supporting information VI)

$$H_{eff} = \sum_{\mathbf{k}}\left[(E_{\mathbf{k}} - \mu' + i\Gamma_{\mathbf{k}})c^+_{\mathbf{k}\uparrow}c_{\mathbf{k}\uparrow} + (E_{\mathbf{k}} - \mu' - i\Gamma_{\mathbf{k}})c^+_{\mathbf{k}\downarrow}c_{\mathbf{k}\downarrow}\right] - g_T\int d^3 r c^+_{\uparrow}(\mathbf{r})c^+_{\downarrow}(\mathbf{r})c_{\downarrow}(\mathbf{r})c_{\uparrow}(\mathbf{r}) \quad (33)$$

where we have defined $c^+_{\mathbf{k}} = \frac{1}{\sqrt{V}}\int d^3 r e^{i\mathbf{k}\cdot\mathbf{r}}c^+(\mathbf{r})$, the total coupling constant coming from both pthe phonon and the dislon gives $g_T = g_{ph} + g_{dis}$. To obtain the superconducting transition temperature with the contribution from both phonons and dislocations, we adopt an auxiliary field approach at a mean-field level[31] (Supporting information VI), finally we obtain that

$$\frac{1}{g_{ph} + g_{dis}} = N(\mu)\int_0^{\omega_D} \frac{\tanh\left(\frac{\xi + \Gamma + i\Gamma}{2T_c}\right) + \tanh\left(\frac{\xi + \Gamma - i\Gamma}{2T_c}\right)}{2(\xi + \Gamma)}d\xi \quad (34)$$

where $\omega_D$ is the Debye frequency, $N(\mu)$ is the density of states at the Fermi level. Since we are interested in electrons near the Fermi level and $\omega_D \ll \mu$, i.e. the decay mainly exists near the Fermi level, we have assumed a constant decay constant

$$\Gamma \sim \Gamma_{k_F} = \frac{\pi m^*}{4\hbar^2 k_F^2 k_{TF}^4} \left(\frac{Ze^2 N}{V}\right)^2 n_{dis} b^2 \left(\frac{1-2\nu}{1-\nu}\right)^2 \qquad (35)$$

which is also consistent with the classical result[32] aside from a phenomenological constant. From Eq. (34), the competition between the classical scattering $\Gamma$ and the quantum interaction $g_{dis}$ is revealed (Figure 2). It can be seen clearly that if the quantum interaction dominates $g_{dis}/g_{ph} \gg \Gamma/\omega_D$, the superconducting transition temperature increases, and vice versa. In particular, the near-linearity of the $T_c = T_c^0$ curve (black-dotted line) indicates the possibility to use a single parameter, the quantum-to-classical ratio

$$\frac{Quantum}{Classical} \sim \frac{g_{dis}/g_{ph}}{\Gamma/\omega_D} \sim \frac{32\pi\hbar^2 k_F^2 \left(\frac{1-\nu}{1-2\nu}\right)^2}{m^* b^2 (\lambda+2\mu)L} \frac{\omega_D}{g_{ph}} = \frac{32\pi\hbar^2 k_F^2 \left(\frac{1-\nu}{1-2\nu}\right)^2}{m^* b^2 (\lambda+2\mu)L} \frac{\omega_D N(\mu)}{[N(\mu)g_{ph}]} \qquad (36)$$

to estimate whether the dislocation will increase or decrease $T_c$. For easier comparison with experiments (Table 2), we note that $g_{ph}$ has the usual energy unit hence Eq. (36) is dimensionless (Supporting information VII). For possible increase of $T_c$, it is preferable for a material to have a small effective mass $m^*$, lower values of rigidity $\lambda$ and $\mu$, and a smaller system size *L*. What is striking is that a combination of many generally-considered independent parameters appearing in Eq. (36), such as electronic properties and material properties, still gives a reasonably comparable quantity of coupling strength.

## Experimental Comparison

Now we are ready to compare the theory using Eqs. (34) and (36) with experiments. In fact, the advantage of Eq. (36) is quite straightforward, since the dislocation density $n_{dis}$ - which is required to compute the absolute magnitude of $g_{dis}$ and $\Gamma$ but is often missing due to the paucity of experimental data- is cancelled out (See Methods). We see that even at this level of approximation, the predicted $T_c$ show excellent quantitative agreement with experiments (Table 2).

In addition to the small influences of dislocation on superconductivity, it is worth mentioning that in some semiconducting monochalgogenide superlattice structures[33, 34] such as PbTe/PbS, introducing dislocations could drive a semiconductor-superconductor phase transition directly when the thin film thickness is small enough. Although some qualitative explanation of the pressure-induced phase transition or dislocation induced flat band[35] were given, no quantitative agreement has been reached to explain the origin of the superconductivity in these systems. Despite the scarcity of data compared with simple metals prevents a full calculation of magnitude of $T_c$ in this system, given its very small effective mass[36], low rigidity[37] and small dislocation grid period, the estimated quantum-to-classical ratio according to Eq. (36) is expected to be exceedingly high, leading to a great enhancement of superconducting transition temperature.

**Conclusions**

In this study, we show that due to dislocation's topological constraint Eq. (4), a quantized dislocation has to obey a different commutation relation Eq. (8). The resulting new quasiparticle, the "dislon", is the first example of 3D system showing behaviors beyond ordinary Bosonic and Fermionic statistics. As a result, the influence of dislocation on electrons is revealed to be composed of two distinct parts, a well-known classical scattering, and a new type of interaction which couples electron pairs through a dislon (Eqs. (24)-(26)). This theory may enable an investigation to enter into a no-man's land, to study the interplay between the crystal dislocation and other components of materials, such as electrons, phonons, magnetic moments, photons. Such studies would help to understand the role of dislocation on material electronic, thermal, magnetic and optical properties etc. at a new level of clarity.

**Methods**

**Experimental Magnitudes**

The simple metal material data are taken from the Landolt–Börnstein database[38] and a few others[39, 40, 41] to ensure consistency, while the data for dislocated superconductors are taken separately[42, 43, 44, 45, 46, 47], which are all compiled in Table 2. Since the dislocation density is unknown for the majority of materials, while the ratio in Eq. (36) does not fix the absolute magnitude of $g_{dis}$, we have chosen a reasonably estimated value $g_{dis} = 0.02 g_{ph}$ to scale all materials (take dislocation density $n_D \sim 10^{12} cm^{-2}$,

$L=10nm$ and other values from Zn as example, $\frac{g_{dis}}{g_{ph}} \sim \left(\frac{N}{V}\right)^2 \frac{n_{dis}}{L} \frac{(4\pi Ze^2)^2}{2k_{TF}^4(\lambda+2\mu)}$). A different choice of parameters in the reasonable range would slightly change the magnitude but not qualitative behavior. Since $g_{dis}$ and $\Gamma$ have different dimensions, for computational purpose we normalize $g_{dis} \to g_{dis}/V$ to match experimental dimensions of $g_{dis}$ so as all coupling strength has energy dimension. For estimation of the Fermi-wavevector using the free-electron model, we use $k_F = \left(\frac{9\pi}{4}\right)^{1/3} \frac{1}{a_0 r_s}$, where the density parameter is defined as $\frac{4\pi}{3} n a_0^3 r_s^3 \equiv 1$.

**Acknowledgements**

ML would thank H. Liu, R. Jaffe, O. Yuzephovich, J. Synder, L. Fu, L. Meroueh and R. Hanus for their helpful discussions. ML, QS, MSD and GC would like to thank support by S$^3$TEC, an Energy Frontier Research Center funded by U.S. Department of Energy (DOE), Office of Basic Energy Sciences (BES) under Award No. DE-SC0001299/DE-FG02-09ER46577.

**Additional information**

Supplementary information is available in the online version of the paper. Reprints and permissions information is available online. Correspondence and requests for materials should be addressed to ML (mingda@mit.edu) or GC (gchen2@mit.edu).

**Competing financial interests**

The authors declare no competing financial interests.

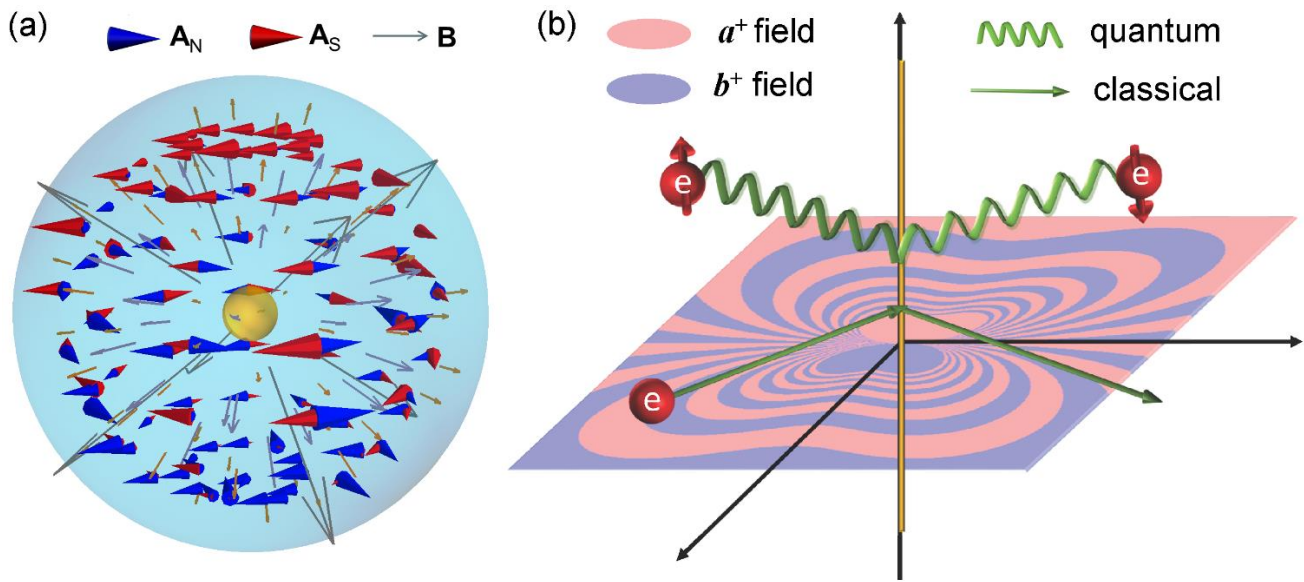

**Figure 1 | Dirac monopole vs Quantized dislocation.** In (a), to decscribe a Dirac monopole (golden sphere), two classical magnetic vector potentials ($A_N$ and $A_S$) have to be implemented due to the non-trivial topology, giving the same magnetic field B (arrows). In (b), to describe a quantized dislocation (golden line), two quantum fields ($a^+$ and $b^+$) are implemented to capture the topology. The color of the field distribution is for illustrative purpose only. The electron-dislocation interactions have two types. The classical interaction (green straight lines with arrows) denotes the momentum-transfer scattering resulting in weakened superconductivity, while another quantum interaction (green wavy lines) leads to an effective attraction between electrons, resulting in enhanced superconductivity.

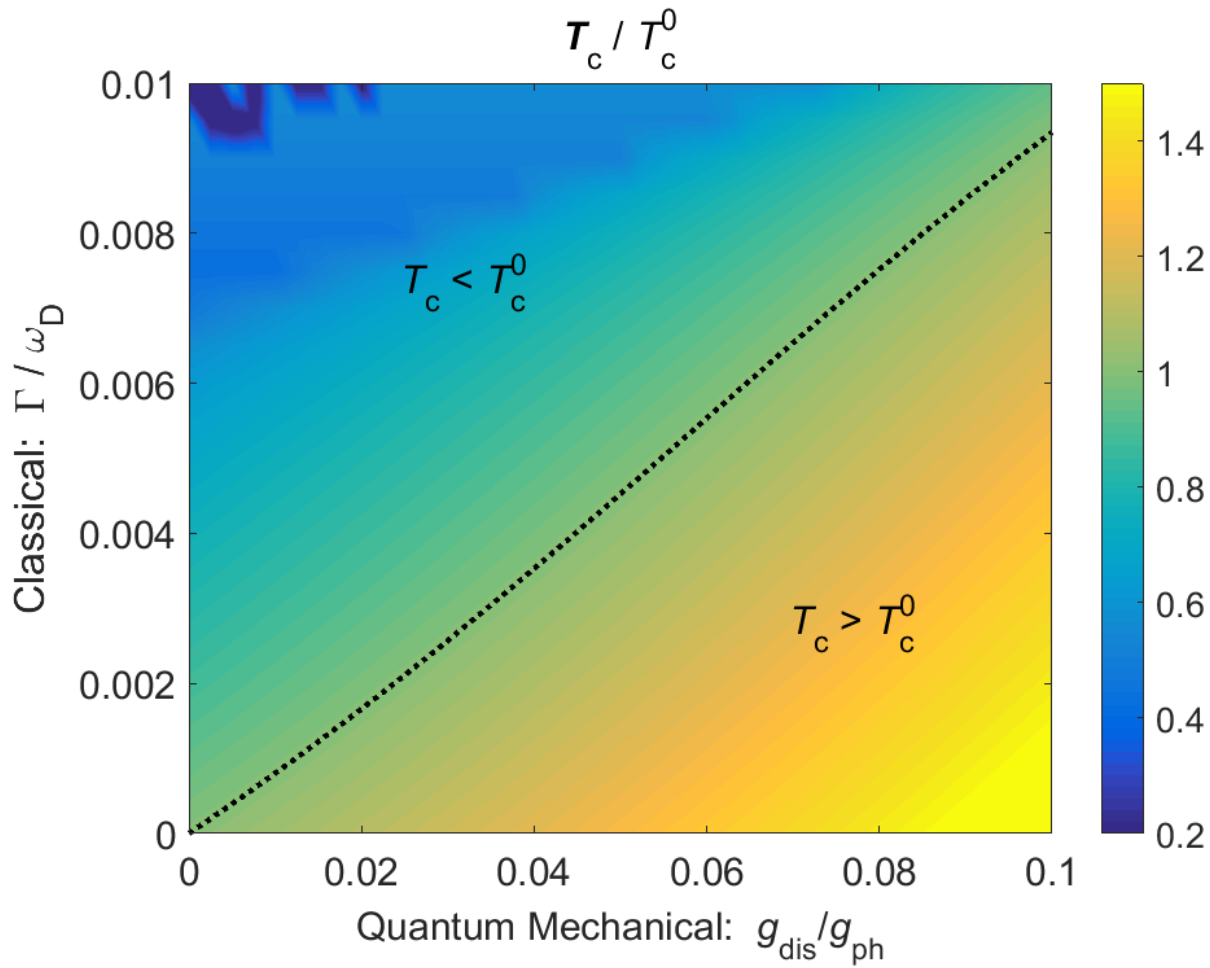

**Figure 2 | Competition between classical and quantum interactions.** There is a clear line (black-dotted) separating the dislocation-enhanced superconductivity ($T_c > T_c^0$) region from dislocation-weakened superconductivity ($T_c < T_c^0$) region.

**Table 1 | Comparison between a phonon and a dislon.** The topological constraint of a dislocation leads to a series of differences compared with a phonon, but still sharing many structural similarities.

| | Phonon | Quantized dislocation (Dislon) |
|---|---|---|
| **Essence** | small displacement **u** | displacement **u** with constraint $\oint d\mathbf{u} = -\mathbf{b}$ |
| **Mode expansion** | $\mathbf{u}(\mathbf{R}) \sim \sum_{\mathbf{k}} u_{\mathbf{k}} \boldsymbol{\varepsilon}_{\mathbf{k}} e^{i\mathbf{k} \cdot \mathbf{R}}$ | $\mathbf{u}(\mathbf{R}) \sim \sum_{\mathbf{k}} u_{\mathbf{k}} \mathbf{F}(\mathbf{k}) e^{i\mathbf{k} \cdot \mathbf{R}}$ |
| **Static limit** | $\lim_{\mathbf{k} \to 0} u_{\mathbf{k}} = 0$ | $\lim_{\kappa \to 0} u_{\mathbf{k}} = 1$ |
| **Electron-interaction** | $H_{e-ph} = \sum_{\mathbf{k'k}\sigma} g_{\mathbf{k}}^{ph} c_{\mathbf{k'+k}\sigma}^{+} c_{\mathbf{k'}\sigma} (A_{\mathbf{k}} + A_{-\mathbf{k}}^{+})$ | $H_{e-dis} = \sum_{\mathbf{k'k}\sigma} g_{\mathbf{k}} c_{\mathbf{k'+k}\sigma}^{+} c_{\mathbf{k'}\sigma} (a_{\mathbf{k}} + a_{-\mathbf{k}}^{+})$ |
| **Algebra** | Bosonic $[A_{\mathbf{k}}, A_{\mathbf{k'}}^{+}] = \delta_{\mathbf{kk'}}$ | Two half-Bosons $[a_{\mathbf{k}}, a_{\mathbf{k'}}^{+}] = \delta_{\mathbf{kk'}} \text{sgn}(\mathbf{k})$ |
| **Coupling Constant with electron** | $g_{\mathbf{k}}^{ph} = \frac{ieV_{\mathbf{k}}}{V}(\mathbf{k} \cdot \boldsymbol{\varepsilon}_{\mathbf{k}}) \sqrt{\frac{\hbar N}{2M\omega_{\mathbf{k}}}}$ | $g_{\mathbf{k}} \equiv \frac{ieN}{VA} V_{\mathbf{k}} [\mathbf{k} \cdot \mathbf{F}(\mathbf{k})] \sqrt{\frac{\hbar N_D}{2m_{\mathbf{k}}\Omega_{\mathbf{k}}}}$ |
| **Effective Electron-electron interaction** | $V_{eff}^{ph}(\mathbf{q}) = -\frac{\hbar\omega_{\mathbf{q}} \left\| g_{\mathbf{q}}^{ph} \right\|^2}{-\omega^2 + \hbar^2\omega_{\mathbf{q}}^2}$ | $V_{eff}(\mathbf{q}) = -\frac{\hbar\Omega_{\mathbf{q}} g_{\mathbf{q}}^{*} g_{\mathbf{q}}}{-\omega^2 + \hbar^2\Omega_{\mathbf{q}}^2}$ |
| **Superconductivity** | $\frac{1}{g_{ph}} = N(\mu) \int_{0}^{\omega_D} \frac{\tanh\left(\frac{\lambda(\xi)}{2T}\right)}{\lambda(\xi)} d\xi$ | $\frac{1}{g_T} = N(\mu') \int_{0}^{\omega_D} \frac{\sum_{s=\pm 1} \tanh\left(\frac{\lambda(\xi) + is\Gamma}{2T}\right)}{2\lambda(\xi)} d\xi$ |

**Table 2 | Comparison between theory and experimental data of superconducting transition temperature for a series of simple-metal dislocated superconductors.** With input of material electronic properties parameters such as the density of state at Fermi level ( $N(\mu)$ ), the electron effective mass $m^*$ and the Fermi wavevector $k_F$ etc., and material mechanical properties including the Poisson ratio $v$, the shear modulus $\mu$, the superconducting transition temperature with the presence of dislocation $T_c$ can be computed and compared with experimental value $[T_c]_{\text{exp}}$. The theory gives a correct trend of all materials.

| | Al | In | Nb | Pb | Sn | Ta | Ti | Tl | V | Zn |
|---|---|---|---|---|---|---|---|---|---|---|
| **Type** | I | I | II | I | I | I | I | I | II | I |
| $N(\mu)g_{ph}$ | 0.175 | 0.29 | 0.29 | 0.36 | 0.245 | 0.25 | 0.15 | 0.27 | 0.24 | 0.165 |
| $N(\mu)$ (eV$^{-1}$) | 0.41 | 0.334 | 0.87 | 0.26 | 0.05 | 0.76 | 0.463 | 0.73 | 1.01 | 0.314 |
| $\omega_D$ (K) | 433 | 112 | 276 | 105 | 199 | 246 | 420 | 78.5 | 399 | 329 |
| $[T_c^0]_{\text{exp}}$ (K) | 1.2 | 3.37 | 9.42 | 7.21 | 3.72 | 4.46 | 0.49 | 2.2 | 5.47 | 0.9 |
| $[T_c]_{\text{exp}}$ (K) | 1.2 | 3.51 | 9.94 | 7.21 | 3.95 | 4.46 | 0.37 | 2.48 | 5.94 | 1.39 |
| $k_F$ (nm$^{-1}$) | 17.5 | 15.0 | 18.9 | 15.7 | 16.2 | 19.5 | 17.5 | 13.5 | 19.5 | 15.7 |
| $m^*/m_e$ | 1.10 | 0.11 | 0.53 | 1.58 | 0.035 | 0.82 | 1.43 | 0.87 | 1.30 | 1.25 |
| $\mu$ (GPa) | 26 | 4 | 38 | 5.6 | 18 | 69 | 44.0 | 2.8 | 47 | 43 |
| $\lambda$ (GPa) | 59 | 32 | 145 | 42 | 46 | 154 | 81 | 41 | 129 | 41 |
| $v$ | 0.35 | 0.45 | 0.40 | 0.44 | 0.36 | 0.34 | 0.32 | 0.45 | 0.37 | 0.25 |
| $Q/CV$ | 14.0 | 234 | 26.7 | 5.68 | 22.8 | 5.93 | 6.88 | 33.9 | 15.2 | 2.57 |

| $[T_c/T_c^0]_{Exp}$ | 1.00 | 1.04 | 1.06 | 1.00 | 1.06 | 1.00 | 0.75 | 1.13 | 1.09 | 1.54 |
|---|---|---|---|---|---|---|---|---|---|---|
| $[T_c/T_c^0]_{Theory}$ | 0.94 | 1.07 | 1.06 | 1.03 | 1.06 | 1.01 | 0.87 | 1.07 | 1.05 | 1.28 |
| **Experiment Trend** | — | ↑ | ↑ | — | ↑ | — | ↓ | ↑ | ↑ | ↑ |
| **Predicted Trend** | ↓ | ↑ | ↑ | ↑ | ↑ | — | ↓ | ↑ | ↑ | ↑ |

# Supporting information of "*Three-dimensional non-Bosonic non-Fermionic quasiparticle through a quantized topological defect of crystal dislocation*"

Mingda Li[1], Qichen Song[1], M. S. Dresselhaus[2] and Gang Chen[1]
[1]Department of Mechanical Engineering, MIT, Cambridge, MA 02139, USA
[2]Department of Physics and Department of Electrical Engineering and Computer Sciences, MIT, Cambridge, MA 02139, USA
Correspondence to: mingda@mit.edu (ML) and gchen2@mit.edu (GC)

## I. The classical static limit

We explicitly compute Eq. (2) in the main text to see why Eq. (4) ensure the displacement of a dislocation. For a screw dislocation along the z-direction, we have $\mathbf{b}=(0\ 0\ b)$ and $n=(0\ 1\ 0)$, hence in the classical limit $\kappa \to 0$, we have

$F_x(\mathbf{s})=0$, $F_y(\mathbf{s})=0$, and $F_z(\mathbf{s})=\dfrac{b}{s^2}\dfrac{k_y}{k_x}$. This gives a displacement $\mathbf{u}_x(\mathbf{R})=\mathbf{u}_y(\mathbf{R})=0$, and most importantly, $\mathbf{u}_z(\mathbf{R})=\dfrac{1}{A}\sum_{\mathbf{s}}e^{i\mathbf{s}\cdot\mathbf{r}}\times\dfrac{1}{s^2}\dfrac{bk_y}{k_x}=\dfrac{b}{4\pi^2}\int\dfrac{k_y}{k_x(k_x^2+k_y^2)}e^{ixk_x+iyk_x}dk_xdk_y=\dfrac{b}{2\pi}\arctan\left(\dfrac{y}{x}\right)$, which is identical with the well-known classical results[1].

For an edge dislocation, we have $\mathbf{b}=(b\ 0\ 0)$, $n=(0\ 1\ 0)$, giving $F_z(\mathbf{s})=0$

$F_x(\mathbf{s})=+\dfrac{b}{k_x s^2}\left(k_y-\dfrac{1}{(1-\nu)}\dfrac{k_x^2 k_y}{s^2}\right)$ and $F_y(\mathbf{s})=+\dfrac{b}{k_x s^2}\left(k_x+-\dfrac{1}{(1-\nu)}\dfrac{k_y^2 k_x}{s^2}\right)$. The corresponding displacements are $\mathbf{u}_z(\mathbf{R})=0$,

$$\mathbf{u}_x(\mathbf{R})=\dfrac{b}{4\pi^2}\int dk_x dk_y \dfrac{e^{ik_x x+ik_y y}}{k_x^2+k_y^2}\left(\dfrac{k_y}{k_x}-\dfrac{1}{(1-\nu)}\dfrac{k_x k_y}{k_x^2+k_y^2}\right)=\dfrac{b}{2\pi}\left[\tan^{-1}\left(\dfrac{y}{x}\right)+\dfrac{1}{2(1-\nu)}\dfrac{xy}{x^2+y^2}\right]$$

$$\mathbf{u}_y(\mathbf{R})=\dfrac{b}{4\pi^2}\int dk_x dk_y \dfrac{e^{ik_x x+ik_y y}}{k_x^2+k_y^2}\left(1-\dfrac{1}{(1-\nu)}\dfrac{k_y^2}{k_x^2+k_y^2}\right)=-\dfrac{b}{2\pi}\left[\dfrac{1-2\nu}{2(1-\nu)}\ln\sqrt{x^2+y^2}+\dfrac{1}{2(1-\nu)}\dfrac{x^2}{x^2+y^2}\right]$$

which all nicely reduced to well-known classical results. In other words, the full definition of the dislocation $\oint d\mathbf{u}=-\mathbf{b}$ is greatly reduced to an equivalent but much simpler condition of Eq. (4) in the main text.

## II. Hamiltonian of the crystal dislocation in its 3D form

The total kinetic energy can be written as

$$T = \frac{\rho}{2}\int \sum_{i=1}^{3} \dot{\mathbf{u}}_i^2(\mathbf{R}) d^3\mathbf{R} = \frac{\rho}{2A^2}\int \sum_{i=1}^{3}\sum_{\mathbf{k}\mathbf{k}'} F_i(\mathbf{k})F_i(\mathbf{k}')e^{i(\mathbf{k}+\mathbf{k}')\cdot\mathbf{R}}\dot{u}_\mathbf{k}\dot{u}_{\mathbf{k}'}d^3\mathbf{R}$$

$$= \frac{\rho}{2L}\sum_{i=1}^{3}\sum_{\mathbf{k}} F_i(\mathbf{k})F_i(-\mathbf{k})\dot{u}_\mathbf{k}\dot{u}_{-\mathbf{k}} = \frac{\rho}{2L}\sum_{\mathbf{k}}|F(\mathbf{k})|^2 \dot{u}_\mathbf{k}\dot{u}_{-\mathbf{k}}$$

(II.1)

Since the distortion tensor can be computed based as

$$u_{ij}(\mathbf{R}) = u_{ji}(\mathbf{R}) = \frac{1}{2}\left(\frac{\partial u_i}{\partial R_j} + \frac{\partial u_j}{\partial R_i}\right) = \frac{1}{A}\sum_{\mathbf{k}\equiv(\mathbf{s},\kappa)} i\frac{k_j F_i(\mathbf{k}) + k_i F_j(\mathbf{k})}{2} e^{i\mathbf{k}\cdot\mathbf{R}} u_\mathbf{k}$$

(II.2)

The total potential can be written as

$$U = \frac{1}{2}\int c_{ijkl} u_{ij} u_{kl} d^3\mathbf{R} = \frac{1}{2}\int \left(\lambda u_{ii} u_{kk} + \mu(u_{ij}u_{ij} + u_{ij}u_{ji})\right) d^3\mathbf{R}$$

$$= -\frac{\lambda}{2}\int \frac{1}{A^2}\sum_{\mathbf{k},\mathbf{k}'}[\mathbf{k}\cdot\mathbf{F}(\mathbf{k})][\mathbf{k}'\cdot\mathbf{F}(\mathbf{k}')]e^{i(\mathbf{k}+\mathbf{k}')\cdot\mathbf{R}}u_\mathbf{k}u_{\mathbf{k}'}d^3\mathbf{R}$$

$$-\frac{\mu}{A^2}\int \sum_{\mathbf{k},\mathbf{k}',i,j}\frac{[k_j F_i(\mathbf{k}) + k_i F_j(\mathbf{k})][k'_j F_i(\mathbf{k}') + k'_i F_j(\mathbf{k}')]}{4}e^{i(\mathbf{k}+\mathbf{k}')\cdot\mathbf{R}}u_\mathbf{k}u_{\mathbf{k}'}d^3\mathbf{R}$$

$$= \frac{(\lambda+\mu)}{2L}\sum_\mathbf{k}[\mathbf{k}\cdot\mathbf{F}(\mathbf{k})]^2 u_\mathbf{k}u_{-\mathbf{k}} + \frac{\mu}{2L}\sum_\mathbf{k} k^2|F(\mathbf{k})|^2 u_\mathbf{k}u_{-\mathbf{k}}$$

(II.3)

Hence the total classical Hamiltonian can be written as Eq. (5) in the main text.

Now for the algebra defined in Eq. (9) in the main text, there is a relevant 2D version

$$\text{sgn}(\mathbf{s}) = \begin{cases} +1, & \text{if } k_x > 0 \\ -1, & \text{if } k_x < 0 \\ \text{sgn}(k_y), & \text{if } k_x = 0 \end{cases}, \quad \text{sgn}(k_y) = \begin{cases} 1, & \text{if } k_y > 0 \\ -1, & \text{if } k_y < 0 \\ 0, & \text{if } k_y = 0 \end{cases}$$

(II.4)

The vector sgn function satisfies a few properties which will prove later to be handy:

**Lemma 1.** $\quad \text{sgn}(-\mathbf{k}) = -\text{sgn}(\mathbf{k}), \text{ for } \forall \mathbf{k}; \; \text{sgn}(\mathbf{k}) = 0 \text{ iff } \mathbf{k} = 0$
$\quad\quad\quad\quad \text{sgn}(-\mathbf{s}) = -\text{sgn}(\mathbf{s}), \text{ for } \forall \mathbf{s}; \; \text{sgn}(\mathbf{s}) = 0 \text{ iff } \mathbf{s} = 0$

where "iff" means "if and only if". The proof is straightforward.

**Lemma 2.** $\{\text{sgn}\,\mathbf{s} \geq 0 | \kappa = 0\} \subset \{\text{sgn}\,\mathbf{k} \geq 0\}$.

Where "$\{\ \}$" is the notation for set. "$X|C$" means set X satisfies condition C.

**Proof:** $\{\text{sgn}\,\mathbf{s} \geq 0 | \kappa = 0\} = \begin{cases} k_x > 0,\ \forall k_y,\ \kappa = 0 \\ k_x = 0,\ k_y > 0,\ \kappa = 0 \\ k_x = 0,\ k_y = 0,\ \kappa = 0 \end{cases} \subset \begin{cases} k_x > 0,\ \forall k_y,\ \forall \kappa \\ k_x = 0,\ k_y > 0,\ \forall \kappa \\ k_x = 0,\ k_y = 0,\ \kappa \geq 0 \end{cases} = \{\text{sgn}\,\mathbf{k} \geq 0\}$

**Lemma 3.** $\{\text{sgn}\,\mathbf{s} \geq 0 | \kappa = 0\} = \{\text{sgn}\,\mathbf{k} \geq 0 | \kappa = 0\}$

**Proof:** Using Lemma 2, $\{\text{sgn}\,\mathbf{s} \geq 0 | \kappa = 0\} = \begin{cases} k_x > 0,\ \forall k_y,\ \kappa = 0 \\ k_x = 0,\ k_y > 0,\ \kappa = 0 \\ k_x = 0,\ k_y = 0,\ \kappa = 0 \end{cases} = \{\text{sgn}\,\mathbf{k} \geq 0 | \kappa = 0\}$.

Now the classical Hamiltonian Eq. (6) in the main text can be simplified with Eqs. (7) and (8) as

$$\begin{aligned} H &= \sum_{\mathbf{k}} \frac{\hbar\Omega(\mathbf{k})}{4}\left(a_{\mathbf{k}}^+ a_{\mathbf{k}} + a_{-\mathbf{k}} a_{-\mathbf{k}}^+ + a_{-\mathbf{k}}^+ a_{-\mathbf{k}} + a_{\mathbf{k}} a_{\mathbf{k}}^+\right) \\ &= \sum_{\mathbf{k}} \hbar\Omega(\mathbf{k}) a_{\mathbf{k}}^+ a_{\mathbf{k}} = \sum_{\mathbf{k} \geq 0} \hbar\Omega(\mathbf{k}) a_{\mathbf{k}}^+ a_{\mathbf{k}} + \sum_{\mathbf{k} < 0} \hbar\Omega(\mathbf{k}) b_{-\mathbf{k}} b_{-\mathbf{k}}^+ \\ &= \sum_{\mathbf{k} \geq 0} \hbar\Omega(\mathbf{k})\left(a_{\mathbf{k}}^+ a_{\mathbf{k}} + \frac{1}{2}\right) + \sum_{\mathbf{k} \geq 0} \hbar\Omega(\mathbf{k})\left(b_{\mathbf{k}}^+ b_{\mathbf{k}} + \frac{1}{2}\right) \end{aligned} \tag{II.5}$$

where we have used the fact that $\Omega(\mathbf{k}=\mathbf{0})=0$, and we use the short-hand notation $\sum_{\mathbf{k}\geq 0}$ standing for summing over $\mathbf{k}$ for all $\text{sgn}\,\mathbf{k} \geq 0$.

## III. Derivation of electron-dislocation interaction Hamiltonian

For electron-dislocation scattering, with the generic interaction Hamiltonian between electron and ions in Eq. (12) of main text, we have

$$\begin{aligned} \sum_{j=1}^{N} \nabla_{\mathbf{R}} V_{ei}(\mathbf{R} - \mathbf{R}_j^0) \cdot \mathbf{u}_j &= \frac{N}{VA} \sum_{\substack{\mathbf{q} \in \text{1BZ}, \mathbf{G} \\ \mathbf{k}=(\mathbf{s},\kappa)}} V_{\mathbf{q}+\mathbf{G}} e^{i(\mathbf{q}+\mathbf{G})\cdot\mathbf{R}} \left[i(\mathbf{q}+\mathbf{G}) \cdot \mathbf{F}(\mathbf{k})\right] \sqrt{\frac{\hbar}{2m_{\mathbf{k}}\Omega_{\mathbf{k}}}} \left(a_{\mathbf{k}} + a_{-\mathbf{k}}^+\right) \delta_{\mathbf{k},\mathbf{q}+\mathbf{G}} \\ &= \frac{N}{VA} \sum_{\mathbf{k}} V_{\mathbf{k}} e^{i\mathbf{k}\cdot\mathbf{R}} \left[i\mathbf{k} \cdot \mathbf{F}(\mathbf{k})\right] \sqrt{\frac{\hbar}{2m_{\mathbf{k}}\Omega_{\mathbf{k}}}} \left(a_{\mathbf{k}} + a_{-\mathbf{k}}^+\right) \end{aligned}$$

Thus the interacting Hamiltonian Eq. (12) in the main text can now be simplified as

$$H_{e-dis} = \int d^3\mathbf{R} \rho_e(\mathbf{R}) \sum_{j=1}^{N} \nabla_\mathbf{R} V_{ei}(\mathbf{R} - \mathbf{R}_j^0) \cdot \mathbf{u}_j$$

$$= \int d^3\mathbf{R} \frac{e}{V} \sum_{\mathbf{k'p}\sigma} e^{-i\mathbf{p}\cdot\mathbf{R}} c^+_{\mathbf{k'+p}\sigma} c_{\mathbf{k'}\sigma} \frac{N}{VA} \sum_\mathbf{k} V_\mathbf{k} e^{i\mathbf{k}\cdot\mathbf{R}} [i\mathbf{k}\cdot\mathbf{F}(\mathbf{k})] \sqrt{\frac{\hbar}{2m_\mathbf{k}\Omega_\mathbf{k}}} (a_\mathbf{k} + a^+_{-\mathbf{k}})$$

$$= \frac{eN}{VA} \sum_{\mathbf{k'}\sigma} c^+_{\mathbf{k'+k}\sigma} c_{\mathbf{k'}\sigma} \sum_\mathbf{k} V_\mathbf{k} [i\mathbf{k}\cdot\mathbf{F}(\mathbf{k})] \sqrt{\frac{\hbar}{2m_\mathbf{k}\Omega_\mathbf{k}}} (a_\mathbf{k} + a^+_{-\mathbf{k}}) \equiv \sum_{\mathbf{k'k}\sigma} g_\mathbf{k} c^+_{\mathbf{k'+k}\sigma} c_{\mathbf{k'}\sigma} (a_\mathbf{k} + a^+_{-\mathbf{k}})$$

## IV. Effective electron action using functional integral approach

To simplify the effective action Eq. (22) in the main text, Fourier transform is first performed,

$$e^{-S_{eff}[\bar{\psi},\psi]} \equiv e^{-S_e[\bar{\psi},\psi]} \int D[\bar{d},d] \prod_{ns} \int \frac{d\bar{k}_{ns}}{2\pi} \frac{dk_{ns}}{2\pi} e^{i\bar{k}_{ns}(d_{ns0}-C_s)} e^{ik_{ns}(\bar{d}_{ns0}-C_s)} e^{-S_{dis}[\bar{d},d] - S_{e-dis}[\bar{\psi},\psi,\bar{d},d]}$$

$$= e^{-S_e[\bar{\psi},\psi]} \times \int D[\bar{d},d] D[\bar{k},k] e^{i \sum_{n,s\geq0}(\bar{k}_{ns}d_{ns0} + k_{ns}\bar{d}_{ns0} - C_s k_{ns} - C_s \bar{k}_{ns})} e^{-S_{dis}[\bar{d},d] - S_{e-dis}[\bar{\psi},\psi,\bar{d},d]}$$

(IV.1)

Where we have defined that functional measure $D[\bar{k},k] \equiv \prod_{ns} \int \frac{d\bar{k}_{ns}}{2\pi} \frac{dk_{ns}}{2\pi}$.

Further integrating over dislon degree of freedom, and using Lemma 3, we have

$$e^{-S_{eff}[\bar{\psi},\psi]} = e^{-S_e[\bar{\psi},\psi]} \int D[\bar{d},d] D[\bar{k},k] \times \exp\left[\sum_{ns\geq0} -iC_s(\bar{k}_{ns} + k_{ns})\right] \times$$

$$\exp\left[-\sum_{n\mathbf{k}\geq0} \bar{d}_{\mathbf{k}n}(-i\omega_n + \hbar\Omega_\mathbf{k})d_{\mathbf{k}n} + \sum_{n\mathbf{k}\geq0}\left(i\bar{k}_{ns}\delta_{\kappa 0} - \frac{g_\mathbf{k}}{\sqrt{\beta/2}}\rho_{n\mathbf{k}}\right)d_{n\mathbf{k}} + \sum_{n\mathbf{k}\geq0}\left(ik_{ns}\delta_{\kappa 0} - \frac{g^*_\mathbf{k}}{\sqrt{\beta/2}}\rho_{-n-\mathbf{k}}\right)\bar{d}_{n\mathbf{k}}\right]$$

$$= e^{-S_e[\bar{\psi},\psi]} \times \exp\left[\sum_{n\mathbf{k}\geq0} \frac{2g^*_\mathbf{k} g_\mathbf{k}}{\beta(-i\omega_n + \hbar\Omega_\mathbf{k})} \rho_{-n-\mathbf{k}} \rho_{n\mathbf{k}}\right] \times \int D[\bar{k},k] \times$$

$$\exp\left[\sum_{ns\geq0} -\frac{\bar{k}_{ns}k_{ns}}{(-i\omega_n + \hbar\Omega_s)} - i\left(\frac{g^*_s \rho_{-n-s}}{\sqrt{\beta/2}(-i\omega_n + \hbar\Omega_s)} + C_s\right)\bar{k}_{ns} - i\left(\frac{g_s \rho_{ns}}{\sqrt{\beta/2}(-i\omega_n + \hbar\Omega_s)} + C_s\right)k_{ns}\right]$$

Now we further integrate over the $[\bar{k},k]$ fields,

$$e^{-S_{eff}[\bar{\psi},\psi]} = e^{-S_e[\bar{\psi},\psi]} \times \exp\left[\sum_{n\mathbf{k}\geq 0} \frac{2g_\mathbf{k}^* g_\mathbf{k}}{\beta(-i\omega_n + \hbar\Omega_\mathbf{k})} \rho_{-n-\mathbf{k}}\rho_{n\mathbf{k}}\right] \times$$

$$\exp\left[\sum_{n\mathbf{s}\geq 0}\left(-\frac{2g_\mathbf{s}^* g_\mathbf{s} \rho_{-n-\mathbf{s}}\rho_{n\mathbf{s}}}{\beta(-i\omega_n + \hbar\Omega_\mathbf{s})} - \frac{C_\mathbf{s} g_\mathbf{s}}{\sqrt{\beta/2}}\rho_{n\mathbf{s}} - \frac{g_\mathbf{s}^* C_\mathbf{s}}{\sqrt{\beta/2}}\rho_{-n-\mathbf{s}} - C_\mathbf{s}^2(-i\omega_n + \hbar\Omega_\mathbf{s})\right)\right]$$

Where the constant terms $C_\mathbf{s}^2(-i\omega_n + \hbar\Omega_\mathbf{s})$ can be neglected. Now we deal with the quartic term $\rho_{-n-\mathbf{k}}\rho_{n\mathbf{k}}$. Using Lemma 3, we have $\sum_{\mathbf{k}\geq 0}\ldots - \sum_{\mathbf{s}\geq 0}\delta_{\kappa 0}\ldots = \sum_{\mathbf{k}\geq 0}\ldots - \sum_{\mathbf{k}\geq 0}\delta_{\kappa 0}\ldots = \sum_{\mathbf{k}\geq 0|\kappa\neq 0}\ldots = \frac{1}{2}\sum_{\mathbf{k}|\kappa\neq 0}\ldots$, and using Lemma 1, and the fact $g_{-\mathbf{s}} = g_\mathbf{s}^*$, $\sum_n = \sum_{-n}$

we have the effective action can further be reduced as

$$e^{-S_{eff}[\bar{\psi},\psi]} = e^{-S_e[\bar{\psi},\psi]} \times \exp\left[\sum_{n\mathbf{k}|\kappa\neq 0} \frac{g_\mathbf{k}^* g_\mathbf{k}}{\beta(-i\omega_n + \hbar\Omega_\mathbf{k})}\rho_{-n-\mathbf{k}}\rho_{n\mathbf{k}}\right]\times\exp\left[-\sum_{n\mathbf{s}}\left(\frac{C_\mathbf{s} g_\mathbf{s}}{\sqrt{2\beta}}\rho_{n\mathbf{s}} + \frac{g_\mathbf{s}^* C_\mathbf{s}}{\sqrt{2\beta}}\rho_{-n-\mathbf{s}}\right)\right]$$

$$= e^{-S_e[\bar{\psi},\psi]} \times \exp\left[\sum_{n\mathbf{k}|\kappa\neq 0} \frac{\hbar\Omega_\mathbf{k} g_\mathbf{k}^* g_\mathbf{k}}{\beta(\omega_n^2 + \hbar^2\Omega_\mathbf{k}^2)}\rho_{-n-\mathbf{k}}\rho_{n\mathbf{k}} - \sum_{\mathbf{k}n\sigma}\sum_{m\mathbf{s}}\frac{C_\mathbf{s}}{\sqrt{2\beta}}\left(\bar{\psi}_{n+m,\mathbf{k}+\mathbf{s}\sigma}g_\mathbf{s} + \bar{\psi}_{n-m,\mathbf{k}-\mathbf{s}\sigma}g_\mathbf{s}^*\right)\psi_{n\mathbf{k}\sigma}\right]$$

Finally, we have the effective action can be written as

$$S_{eff}[\bar{\psi},\psi] = S_e[\bar{\psi},\psi] - \sum_{n\mathbf{k}|\kappa\neq 0}\frac{\hbar\Omega_\mathbf{k} g_\mathbf{k}^* g_\mathbf{k}}{\beta(\omega_n^2 + \hbar^2\Omega_\mathbf{k}^2)}\rho_{-n-\mathbf{k}}\rho_{n\mathbf{k}} + \sum_{n\mathbf{s}}\left(\frac{C_\mathbf{s} g_\mathbf{s}}{\sqrt{2\beta}}\rho_{n\mathbf{s}} + \frac{g_\mathbf{s}^* C_\mathbf{s}}{\sqrt{2\beta}}\rho_{-n-\mathbf{s}}\right) \quad \text{(IV.2)}$$

This is equivalent with Eq. (23) in the main text by noticing that $\rho_{n\mathbf{s}} \equiv \sum_{\mathbf{k}'m\sigma}\bar{\psi}_{m+n,\mathbf{k}'+\mathbf{s}\sigma}\psi_{m\mathbf{k}'\sigma}$ where $\mathbf{s}$ is the in-plane 2D momentum.

## V.  Electron-classical dislocation interaction

For an edge dislocation, $\mathbf{b} = (b\ 0\ 0)$, $\mathbf{b}\cdot\mathbf{s} = bk_x$, $\mathbf{n}\cdot\mathbf{s} = s\sin\theta$ and hence we have the magnitude of the classical scattering amplitude based on Eq. (25) in the main text as

$$A_\mathbf{s}^{Edge} = \frac{N}{2VL}eV_\mathbf{s}\left(\frac{1-2\nu}{1-\nu}\right)\frac{b\sin\theta}{\sqrt{k_x^2 + k_y^2}} \quad \text{(V.1)}$$

where $\theta$ is the angle between Burgers vector and dislocation line direction. Now performing Fourier transform, Assuming $V_s$ is independent of $\mathbf{s}$ due to screening effect, we have (Assuming $\mathbf{r} \neq 0$)

$$
\begin{aligned}
A^{Edge}(\mathbf{r}) &= \int A_\mathbf{s}^{Edge} e^{i\mathbf{s}\cdot\mathbf{r}} \frac{d^2\mathbf{s}}{4\pi^2} = \frac{N}{2VL}\frac{4\pi Ze^2}{k_{TF}^2}\sin\theta\left(\frac{1-2\nu}{1-\nu}\right)\frac{b}{4\pi^2}\int_{-\infty}^{+\infty} dk_y e^{ik_y y} \int_{-\infty}^{+\infty} dk_x \frac{e^{ik_x x}}{\sqrt{k_x^2+k_y^2}} \\
&= \frac{N}{2VL}\sqrt{n_{dis}}\frac{4\pi Ze^2}{k_{TF}^2}\sin\theta\left(\frac{1-2\nu}{1-\nu}\right)\frac{b}{4\pi^2}\underbrace{\int_{-\infty}^{+\infty} dk_y\, 2K_0(|xk_y|) e^{ik_y y}}_{} \\
&= \frac{N}{2VL}\sqrt{n_{dis}}\frac{4\pi Ze^2}{k_{TF}^2}\times\frac{b\sin\theta}{4\pi^2}\left(\frac{1-2\nu}{1-\nu}\right)\underbrace{\frac{2\pi}{\sqrt{x^2+y^2}}}_{} = \frac{N\sqrt{n_{dis}}}{VL}\frac{2\pi Ze^2}{k_{TF}^2}\times\frac{b}{2\pi}\left(\frac{1-2\nu}{1-\nu}\right)\frac{\sin\theta}{r} \\
&\propto \frac{b}{2\pi}\left(\frac{1-2\nu}{1-\nu}\right)\frac{\sin\theta}{r}
\end{aligned}
\quad (V.2)
$$

which shows excellent agreement with the well-known classical electron-dislocation scattering.

For the elastic scattering rate, we have

$$
\begin{aligned}
\Gamma_\mathbf{k} &= \frac{\pi}{2}\sum_\mathbf{s}\left|\langle\mathbf{k}+\mathbf{s}|\hat{A}_c|\mathbf{k}\rangle\right|^2\delta(\varepsilon_{\mathbf{k}+\mathbf{s}}-\varepsilon_\mathbf{k}) = \frac{\pi}{\hbar}\sum_\mathbf{s}|A_\mathbf{s}|^2\delta(\varepsilon_{\mathbf{k}+\mathbf{s}}-\varepsilon_\mathbf{k}) = \frac{L^2}{4\pi\hbar}\int|A_\mathbf{s}|^2\delta(\varepsilon_{\mathbf{k}+\mathbf{s}}-\varepsilon_\mathbf{k})d^2\mathbf{s} \\
&= \frac{1}{8\pi}\left(\frac{eN}{2V}\right)^2 n_{dis}b^2\left(\frac{1-2\nu}{1-\nu}\right)^2\int d\theta_s \int_0^{k_F}|V_\mathbf{s}|^2\sin^2\theta_s\delta(\varepsilon_{\mathbf{k}+\mathbf{s}}-\varepsilon_\mathbf{k})\frac{ds}{s} \\
&= \frac{1}{8\pi}\left(\frac{eN}{2V}\right)^2 n_{dis}b^2\left(\frac{1-2\nu}{1-\nu}\right)^2\int d\theta_s \int_0^{k_F}\frac{(4\pi Ze)^2}{(s^2+k_{TF}^2)^2}\sin^2\theta_s\,\frac{m^*}{\hbar^2 k}\delta(s+2k\cos\theta_{ks})\frac{ds}{s} \\
&= -\frac{\pi m^*}{4\hbar^2 k^2}\left(\frac{Ze^2 N}{V}\right)^2 n_{dis}b^2\left(\frac{1-2\nu}{1-\nu}\right)^2\int_0^{2\pi}\frac{\sin^2\theta_s}{(4k^2\cos^2\theta_{ks}+k_{TF}^2)^2\cos\theta_{ks}}d\theta_s
\end{aligned}
\quad (V.3)
$$

where $\theta_s$ is the angle of $\mathbf{s}$ with $\mathbf{x}$ axis, and $\theta_{ks}$ is the angle between $\mathbf{k}$ and $\mathbf{s}$, and the energy conservation can be written as $\delta(\varepsilon_{\mathbf{k}+\mathbf{s}}-\varepsilon_\mathbf{k}) = \frac{m^*}{\hbar^2 k}\delta(s+2k\cos\theta_{ks})$. Now if we assume a long wavelength limit $k \ll k_{TF}$ and neglected the complex angle-dependent integration which is of order of magnitude $\sim O(1)$, the relaxation rate can finally be written as Eq. (29) in the main text.

## VI.　Electron- quantum dislocation interaction: superconductivity

From Eq. (31) in the main text, performing Fourier transform of operator $c_{\mathbf{k}}^{+} = \frac{1}{\sqrt{V}}\int d^3r e^{i\mathbf{k}\cdot\mathbf{r}} c^{+}(\mathbf{r})$, we have

$$\sum_{\mathbf{qkk'}} c_{\mathbf{k+q}\uparrow}^{+} c_{-\mathbf{k}\downarrow}^{+} c_{-\mathbf{k'+q}\downarrow} c_{\mathbf{k'}\uparrow} = \frac{1}{V^2}\int d^3r\, d^3\mathbf{r}_1 d^3\mathbf{r}_2 d^3\mathbf{r}_3 c_{\uparrow}^{+}(\mathbf{r}) c_{\downarrow}^{+}(\mathbf{r}_1) c_{\downarrow}^{+}(\mathbf{r}_2) c_{\uparrow}^{+}(\mathbf{r}_3) \sum_{\mathbf{k}} e^{i\mathbf{k}\cdot(\mathbf{r}-\mathbf{r}_1)} \sum_{\mathbf{q}} e^{i\mathbf{q}\cdot(\mathbf{r}-\mathbf{r}_2)} \sum_{\mathbf{k'}} e^{i\mathbf{k'}\cdot(\mathbf{r}_2-\mathbf{r}_3)}$$

$$= \frac{1}{V^2}\int d^3r\, d^3\mathbf{r}_1 d^3\mathbf{r}_2 d^3\mathbf{r}_3 c_{\uparrow}^{+}(\mathbf{r}) c_{\downarrow}^{+}(\mathbf{r}_1) c_{\downarrow}^{+}(\mathbf{r}_2) c_{\uparrow}^{+}(\mathbf{r}_3) V\delta(\mathbf{r}-\mathbf{r}_1) V\delta(\mathbf{r}-\mathbf{r}_2) V\delta(\mathbf{r}_2-\mathbf{r}_3)$$

$$= V\int d^3r\, c_{\uparrow}^{+}(\mathbf{r}) c_{\downarrow}^{+}(\mathbf{r}) c_{\downarrow}^{+}(\mathbf{r}) c_{\uparrow}^{+}(\mathbf{r})$$

which recovers the Eq. (33) in the main text.

Quantitatively, to obtain the superconducting transition temperature with the presence of dislocations, we adopt the auxiliary field method by doing the Hubbard-Stratonovich transformation,

$$Z = \int D[\bar{\psi},\psi]\exp\left[-\sum_{\mathbf{k}n}\bar{\psi}_{n\mathbf{k}\uparrow}(-ip_n + E_{\mathbf{k}} - \mu' + i\Gamma_{\mathbf{k}})\psi_{n\mathbf{k}\uparrow} - \sum_{\mathbf{k}n}\bar{\psi}_{n\mathbf{k}\downarrow}(-ip_n + E_{\mathbf{k}} - \mu' - i\Gamma_{\mathbf{k}})\psi_{n\mathbf{k}\downarrow}\right]$$

$$\times\exp\left(+g_T\int_0^\beta d\tau\int d^3r\,\bar{\psi}_{\uparrow}(\mathbf{r})\bar{\psi}_{\downarrow}(\mathbf{r})\psi_{\downarrow}(\mathbf{r})\psi_{\uparrow}(\mathbf{r})\right)$$

$$= \int D[\bar{\psi},\psi]\exp\left[-\sum_{\mathbf{k}n}\bar{\psi}_{n\mathbf{k}\uparrow}(-ip_n + E_{\mathbf{k}} - \mu' + i\Gamma_{\mathbf{k}})\psi_{n\mathbf{k}\uparrow} - \sum_{\mathbf{k}n}\bar{\psi}_{n\mathbf{k}\downarrow}(-ip_n + E_{\mathbf{k}} - \mu' - i\Gamma_{\mathbf{k}})\psi_{n\mathbf{k}\downarrow}\right]$$

$$\times \int D[\bar{\Delta},\Delta]\exp\left[-\frac{1}{g_T}\int_0^\beta d\tau\int d^3r\,\bar{\Delta}(\mathbf{r})\Delta(\mathbf{r}) + \int_0^\beta d\tau\int d^3r\,\bar{\Delta}(\mathbf{r})\psi_{\downarrow}(\mathbf{r})\psi_{\uparrow}(\mathbf{r}) + \int_0^\beta d\tau\int d^3r\,\bar{\Delta}(\mathbf{r})\psi_{\downarrow}(\mathbf{r})\psi_{\uparrow}(\mathbf{r})\right]$$

$$= \int D[\bar{\Delta},\Delta]\exp\left[-\frac{1}{g_T}\int_0^\beta d\tau\int d^3r\,\bar{\Delta}(\mathbf{r})\Delta(\mathbf{r})\right]\times$$

$$D[\bar{\psi},\psi]\exp\left[-\int_0^\beta d\tau d^3r\begin{pmatrix}\bar{\psi}_{\uparrow} & \psi_{\downarrow}\end{pmatrix}\begin{pmatrix}\partial_\tau + E - \mu' + i\Gamma & \Delta \\ \bar{\Delta} & \partial_\tau - E + \mu' + i\Gamma\end{pmatrix}\begin{pmatrix}\psi_{\uparrow} \\ \bar{\psi}_{\downarrow}\end{pmatrix}\right]$$

The effective action can now be written as

$$S[\bar{\Delta},\Delta] = \frac{1}{g_T}\int_0^\beta d\tau\int d^3r\,\bar{\Delta}(\mathbf{r})\Delta(\mathbf{r}) - \text{Tr}\ln\beta\begin{bmatrix}\partial_\tau + E - \mu' + i\Gamma & \Delta \\ \bar{\Delta} & \partial_\tau - E + \mu' + i\Gamma\end{bmatrix} \qquad (\text{VI.1})$$

On the mean-field level, assuming a spatial-independent $\Delta$ field $\Delta(\mathbf{r}) \equiv \Delta$, we have

$$\frac{\delta S[\bar{\Delta},\Delta]}{\delta \Delta}=0 \Rightarrow \frac{1}{g_T}\bar{\Delta}=\frac{1}{\beta V}\sum_{n\mathbf{k}}\begin{pmatrix} -ip_n+E_\mathbf{k}-\mu'+i\Gamma_\mathbf{k} & \Delta \\ \bar{\Delta} & -ip_n-E_\mathbf{k}+\mu'+i\Gamma_\mathbf{k} \end{pmatrix}^{-1}\begin{pmatrix} 0 & 1 \\ 0 & 0 \end{pmatrix}$$

$$\frac{1}{g_T}\bar{\Delta}=-\frac{1}{\beta V}\sum_{n\mathbf{k}}\frac{-\bar{\Delta}}{(E_\mathbf{k}-\mu')^2+(p_n-\Gamma_\mathbf{k})^2+|\Delta|^2} \Rightarrow \frac{1}{g_T}=\frac{1}{\beta V}\sum_{n\mathbf{k}}\frac{1}{(E_\mathbf{k}-\mu')^2+(p_n-\Gamma_\mathbf{k})^2+|\Delta|^2}$$

(VI.2)

To compute the frequency summation in Eq. (VI.2), we use complex integration method. Defining an contour integral in the complex plane $I=-\frac{1}{2\pi i}\oint dz \frac{n_F(z)}{(z-i\Gamma_\mathbf{k})^2-(E_\mathbf{k}-\mu')^2-|\Delta|^2}$, where $n_F$ is the Fermi occupation function. The integrand has poles at $z=ip_n$ and $z=i\Gamma_\mathbf{k}\pm\sqrt{\varepsilon_\mathbf{k}^2+|\Delta|^2}$ in the entire complex plane. Now using the residue theorem and using the contour integral over the whole plane, we have

$$I=0=-\sum_n \frac{1}{(ip_n-i\Gamma_\mathbf{k})^2-(E_\mathbf{k}-\mu')^2-|\Delta|^2}\times\left(-\frac{1}{\beta}\right)-\frac{n_F(i\Gamma_\mathbf{k}+\sqrt{(E_\mathbf{k}-\mu')^2+|\Delta|^2})}{2\sqrt{(E_\mathbf{k}-\mu')^2+|\Delta|^2}}+\frac{n_F(i\Gamma_\mathbf{k}-\sqrt{(E_\mathbf{k}-\mu')^2+|\Delta|^2})}{2\sqrt{(E_\mathbf{k}-\mu')^2+|\Delta|^2}}$$

Substituting back to Eq. (VI.2), we have

$$\frac{1}{g_T}=\frac{1}{V}\sum_\mathbf{k}\frac{n_F(i\Gamma_\mathbf{k}-\lambda_\mathbf{k})}{2\lambda_\mathbf{k}}-\frac{n_F(i\Gamma_\mathbf{k}+\lambda_\mathbf{k})}{2\lambda_\mathbf{k}} \qquad (VI.3)$$

Where we have defined the quasiparticle excitation energy $\lambda_\mathbf{k}=\sqrt{(E_\mathbf{k}-\mu')^2+|\Delta|^2}>0$. Further computing ($\lambda(\xi)=\sqrt{(\xi-\mu')^2+|\Delta|^2}>0$), we have

$$\frac{1}{g_T}=\frac{1}{g_{ph}+g_{dis}}=\frac{1}{V}\sum_\mathbf{k}\frac{1}{2\sqrt{\lambda_\mathbf{k}^2+\Gamma_\mathbf{k}^2}}[n_F(i\Gamma_\mathbf{k}-\lambda_\mathbf{k})-n_F(i\Gamma_\mathbf{k}+\lambda_\mathbf{k})]$$

$$=N(\mu)\int_0^{\omega_D}\frac{n_F(i\Gamma-\lambda(\xi))-n_F(i\Gamma+\lambda(\xi))}{\lambda(\xi)}d\xi \qquad (VI.4)$$

$$=N(\mu')\int_0^{\omega_D}\frac{\tanh\left(\frac{\lambda(\xi)+i\Gamma}{2T}\right)+\tanh\left(\frac{\lambda(\xi)-i\Gamma}{2T}\right)}{2\lambda(\xi)}d\xi$$

where we have defined the density of states at the Fermi level as $N(\mu')$. Written in this way, Eq. (VI.4) resembles very much the original BCS gap equation,

$$\frac{1}{g_{ph}} = N(\mu) \int_0^{\omega_D} \frac{\tanh\left(\frac{\lambda(\xi)}{2T}\right)}{\lambda(\xi)} d\xi \tag{VI.5}$$

The meaning of $N(\mu)$ means it is taken from the bare electron spectra $\varepsilon_k$ prior to renormalization, with a different density of states. Noticing that at $T=T_c$, the superconducting energy gap $|\Delta|=0$, we obtain the corresponding $T_c$ equation which is the Eq. (34) in the main text.

## VII. Dimensions of all relevant parameters

$[H] = [T] = [U] = M^{+1}L^2T^{-2}$, $[F(\mathbf{k})] = L^3$, $[\mathbf{u}_i(\mathbf{R})] = L$, $[\mathbf{u}_\mathbf{k}] = 1$, $[\lambda] = [\mu] = M^{+1}L^{-1}T^{-2}$,

$[\rho] = M^{+1}L^{-3}$, $[T(\mathbf{k})] = M^{+1}L^3$, $[U(\mathbf{k})] = M^{+1}L^3T^{-2}$, $[m_\mathbf{k}] = M^{+1}L^2$, $[\Omega_\mathbf{k}] = T^{-1}$, $[\hbar] = M^{+1}L^{+2}T^{-1}$,

$[Z_\mathbf{k}] \equiv 1$, $[e] = M^{1/2}L^{3/2}T^{-1}$ (**static Coulomb**), $[V_\mathbf{q}] = M^{+1/2}L^{+7/2}T^{-1}$, $[eV_\mathbf{q}] = M^{+1}L^{+5}T^{-2}$,

$[\rho_e(\mathbf{r})] = M^{1/2}L^{-3/2}T^{-1}$, $[c(\mathbf{r})] = L^{-3/2}$, $[g_T] = [g_{ph}] = [g_{dis}] = M^{+1}L^5T^{-2}$, $[\Delta(\mathbf{r})] = M^{+1}L^2T^{-2} = [H]$,

$[\psi_{n\mathbf{p}}] = [\chi_{n\mathbf{p}}] = [d_{n\mathbf{p}}] = [C_s] = M^{-1/2}L^{-1}T^{+1} = [H^{-1/2}]$, $[\beta] = [\rho_{n\mathbf{k}}] = M^{-1}L^{-2}T^{+2} = [H^{-1}]$,

$[g_\mathbf{k}] = M^{+1}L^2T^{-2} = [H]$, $[T_c] = M^{+1}L^2T^{-2} = [H]$, $[N(E_F)] = M^{-1}L^{-5}T^{+2} = [g_{ph}]^{-1}$.

However, the experimental magnitude of $g_{ph}$ is usually given in energy units, hence to make the quantum-to-classical ratio comparable, we express all the magnitudes in the unit of energy as: $[g_{dis}] = [g_{ph}] = [\Gamma] = [\omega_D] = M^{+1}L^2T^{-2}$ by dividing both the numerator and denominator by the volume *V*.